\documentclass[twocolumn,english,aps,prl,superscriptaddress,amsmath,amssymb,showpacs]{revtex4-1}
\usepackage[T1]{fontenc}
\usepackage[latin9]{inputenc}
\usepackage{amstext}
\usepackage{graphicx}

\makeatletter
\usepackage{babel}

\makeatother

\usepackage{babel}
\begin{document}

\title{Tunneling chemical exchange reaction $\textrm{D}+\textrm{HD}\rightarrow\textrm{D}_{2}+\textrm{H}$
in solid HD and D$_{2}$ at temperatures below 1$\,$K}

\date{\today}

\author{S. Sheludiakov}

\email{seshel@utu.fi}

\author{J. Ahokas}

\author{J. Järvinen}

\affiliation{Wihuri Physical Laboratory, Department of Physics and Astronomy,
University of Turku, 20014 Turku, Finland}

\author{D. Zvezdov}

\affiliation{Wihuri Physical Laboratory, Department of Physics and Astronomy,
University of Turku, 20014 Turku, Finland}

\affiliation{Institute of Physics, Kazan Federal University, 18 Kremlyovskaya
St., Kazan 42008, Republic of Tatarstan, Russian Federation}

\author{L. Lehtonen}

\author{O. Vainio}

\author{S. Vasiliev}

\affiliation{Wihuri Physical Laboratory, Department of Physics and Astronomy,
University of Turku, 20014 Turku, Finland}

\author{D. M. Lee}

\author{V. V. Khmelenko}

\affiliation{Institute for Quantum Science and Engineering, Department of Physics
and Astronomy, Texas A$\&$M University, College Station, TX, 77843,
USA}
\begin{abstract}
We report on a study of the exchange tunneling reaction D+HD$\rightarrow$D$_{2}$+H
in a pure solid HD matrix and in a D$_{2}$ matrix with 0.23$\%$
HD admixture at temperatures between 130$\,$mK and 1.5$\,$K. We
found that the exchange reaction rates, $k_{exHD}\sim3\times10^{-27}$cm$^{3}$s$^{-1}$
in the pure HD matrix, and $k_{exD_{2}}=9(4)\times10^{-28}$cm$^{-3}$
in the D$_{2}$ matrix are nearly independent of temperature within
this range. This confirms quantum tunnelling nature of these reactions,
and their ability to proceed at temperatures down to absolute zero.
Based on these observations we concluded that exchange tunneling reaction
H+H$_{2}\rightarrow$H$_{2}$+H should also proceed in a H$_{2}$
matrix at the lowest temperatures. On contrary, the recombination
of H atoms in solid H$_{2}$ and D atoms in solid D$_{2}$ is substantially
suppressed at the lowest temperatures as a result of increasing of
violation for resonance tunneling of atoms when they approach each
other. 
\end{abstract}
\maketitle
\section{Introduction}
Solid hydrogen and deuterium form a special class of quantum solids,
where due to a small mass and weak intermolecular interactions, the
effects of quantum tunneling play an important role. Light atomic
impurities like H and D stabilized in matrices of the solid hydrogens
are able to migrate from one lattice site to another by a repetition
of tunneling reactions\cite{Kumada}: 
\begin{equation}
\textrm{H}+\textrm{H}_{2}\rightarrow\textrm{H}_{2}+\textrm{H}.\label{eq:H_H2_diff}
\end{equation}
\begin{equation}
\textrm{D}+\textrm{D}_{2}\rightarrow\textrm{D}_{2}+\textrm{D}\label{eq:D_D2_diff}
\end{equation}
A clear consequence of this is that a recombination of H and D occurs
when two atoms encounter each other in neighboring lattice sites.
Tunneling reactions also take place in $\textrm{H}_{2}-\textrm{D}_{2}$
mixtures and involve both hydrogen isotopes: 
\begin{equation}
\textrm{D}+\textrm{H}_{2}\rightarrow\textrm{HD}+\textrm{H}.\label{eq:D_H2_react}
\end{equation}
\begin{equation}
\textrm{D}+\textrm{HD}\rightarrow\textrm{D}_{2}+\textrm{H}\label{eq:D_HD_react}
\end{equation}
A spectacular production of H atoms occurs as D atoms combine with
H$_{2}$ in reaction (\ref{eq:D_H2_react}) and HD in reaction (\ref{eq:D_HD_react}).
Reactions (\ref{eq:D_HD_react}) and (\ref{eq:D_H2_react}) are well-known
in the gas phase at higher temperatures and were intensively studied
in H$_{2}$-D$_{2}$ solids \cite{Gordon83,Ivliev83,Tsuruta83} at
temperatures down to $\sim$1 K. The difference in the zero-point
energies between the products and reactants (about 400~K) results
in a preferential creation of H atoms while the reverse reactions
are endothermic and do not proceed at low enough temperatures. The
rates for the tunneling reactions (\ref{eq:D_H2_react}) and (\ref{eq:D_HD_react})
were theoretically calculated by Takayanagi et al. \cite{Takayanagi90}
and Hancock et al. \cite{Hancock89}. The rate of the reaction (\ref{eq:D_HD_react})
was measured experimentally in the temperature range 1.9-6.5~K in
a series of works by Miyazaki and Kumada \cite{Lee87,Kumada96}. Reaction
(\ref{eq:D_H2_react}) proceeds two orders of magnitude faster andits
rate was measured only recently by Kumada \cite{Kumada06}.

The main signatures of a tunneling reaction are weak dependence of
its rate on temperature and a large isotopic effect. For a single
atomic impurity in a perfect crystal tunnelling motion due to the
reactions (\ref{eq:H_H2_diff}) and \ref{eq:D_D2_diff} may be very
fast because of resonance in the energies of the initial and final
states, since all locations for the atoms in the lattice are equivalent
to each other. Kagan et al. \cite{KaganLeggettBook92} pointed out
that crystal defects and irregularities violate the condition of resonant
tunneling of H atoms in H$_{2}$ matrix which affects the rate of
a tunneling reaction. Kumada et al. \cite{Kumada02} showed that the
ortho-molecules as 2nd and 3rd nearest neighbors significantly decrease
the rate of the atomic hydrogen tunneling reaction in a para-H$_{2}$
matrix. According to Kagan, tunneling in irregular solids may be stimulated
by phonons, which force tunneling rates to be proportional to $T$
for a single-phonon direct process or to $T^{n=5-9}$ for a two-phonon
Raman process \cite{KaganLeggettBook92}. It was found experimentally
that the recombination rate of H atoms in solid H$_{2}$ follows a
linear dependence on temperature in the temperature range 1.2-4K \cite{Ivliev82},
while the rate of the exchange reaction eq. (\ref{eq:D_HD_react})
remained nearly temperature independent within experimental errors
\cite{Kumada96}. Significant divergence from the linear law was observed
for the H recombination rates in solid H$_{2}$ upon reducing temperature
below 1~K where the H atom recombination decreased by more than two
orders of magnitude for samples cooled down to 150~mK \cite{HinH2_06,Ahokas10}.\\
 Since in solid D$_{2}$ with a small H$_{2}$ or HD admixture atomic
impurities need to approach a H$_{2}$ or HD molecule before taking
part in reactions (\ref{eq:D_H2_react}) and (\ref{eq:D_HD_react}),
these reactions as well as the recombination of atoms are two stage
processes: 1) diffusion due to tunnelling exchange followed by 2)
the reaction itself. The matrix of pure HD represents a different
case, since the atomic impurities in the lattice are always surrounded
by 12 HD neighbours. The diffusion process is not needed for the reaction
(\ref{eq:D_HD_react}) in a solid HD matrix.\\
 In the present work we report on an experimental study of the isotopic
exchange reaction corresponding to eq. (\ref{eq:D_HD_react}) in a
HD matrix in a temperature range 0.13-1.5~K. An opposite limit where
D atoms need to diffuse in order to encounter a HD molecule to react
with was studied in a solid mixture of D$_{2}$:0.23\% HD. Our cryogenic
system is based on a dilution refrigerator, which allows continuous
measurements of the reaction kinetics for experiment durations of
the order of weeks and months, substantially longer than in any previous
work. This makes possible measurements of very slow chemical processes
with very small concentrations of reagents, not accessible before.
Our data for the rate of the reaction (\ref{eq:D_HD_react}) at T=0.13-1.5$\,$K
in a pure HD matrix agree well with previous works \cite{Lee87,Kumada96}
and confirm a very weak temperature dependence in this range. We found
that the reaction (\ref{eq:D_HD_react}) in the D$_{2}$ matrix is
also temperature independent, which indicates that the non-resonant
diffusion due to the exchange reaction (\ref{eq:D_D2_diff}) is not
a limiting stage. The rate of the reaction (\ref{eq:D_HD_react})
in the D$_{2}$ matrix proceeds about 3 times slower than in a pure
HD matrix. At temperature of 150 mK we have not observed any recombination
of D atoms in the D$_{2}$ matrix, while the exchange reaction (\ref{eq:D_HD_react})
was well visible. Since the recombination also involves the diffusive
process, we anticipate that the atomic motion due to the reaction
(\ref{eq:D_D2_diff}) is slowed down substantially when the D atoms
approach each other. Based on that hypothesis we concluded that in
the previous work the exchange tunneling chemical reaction H+H$_{2}\rightarrow$H$_{2}$+H
also takes place in solid molecular hydrogen at 150$\,$mK while the
atomic recombination is suppressed by non-resonant tunnelling when
two H atoms apporoach each other\cite{HinH2_06}.

\section{Experimental setup}

\begin{figure}
\includegraphics[width=1\columnwidth]{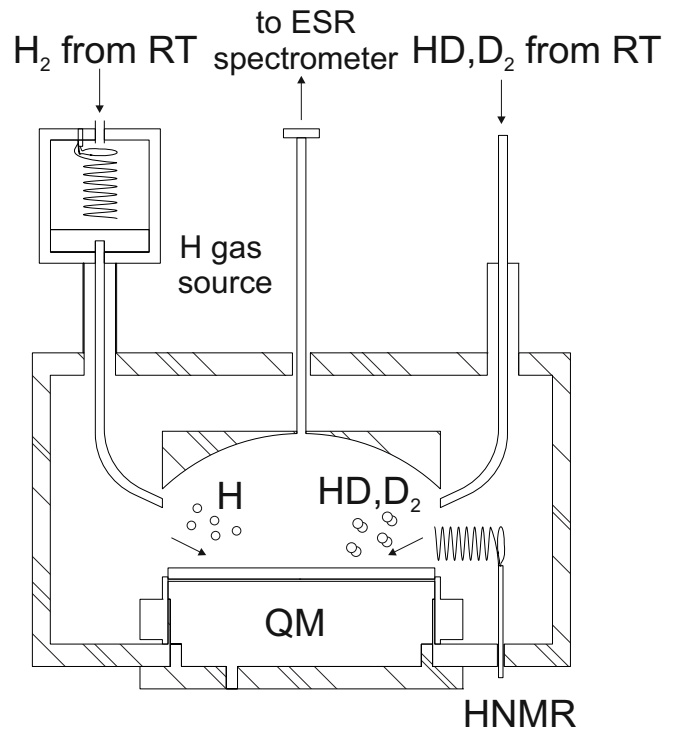}

\protect\protect\caption{Schematic of the sample cell. QM: Quartz crystal microbalance (the
bottom mirror of the Fabry-Perot resonator), HNMR: 910~MHz helical
resonator for RF discharge and H NMR.\label{fig:Sample-cell}}
\end{figure}

Experiments were carried out in the sample cell (SC) shown in Fig.
\ref{fig:Sample-cell} \cite{Cellpaper}. The SC is located in the
center of a 4.6~T superconducting magnet and is cooled by a dilution
refrigerator down to about 100~mK. The main diagnostic tools in our
experiments are an electron spin resonance (ESR) spectrometer employing
a Fabry-Perot resonator operating at 128~GHz frequency \cite{Vasilyev04}
and a quartz microbalance (QM) to measure the sample thickness. ESR
enables measurements of H and D concentrations and hyperfine state
populations. Absolute atomic concentration calibration is obtained
from the dipolar broadening of the ESR spectra \cite{Ahokas10} or
calorimetrically by recording gas phase hydrogen recombination via
measuring heat released due to reactions \cite{Ahokas10,Cellpaper}.
Solid films of hydrogen isotopes were deposited on the flat ESR resonator
mirror which also serves as the QM electrode.

The samples were condensed directly from a room temperature reservoir
through an electrically heated capillary (Fig.\ref{fig:Sample-cell}).
The capillary end is directed at the QM which is kept at temperature
of 0.5-1~K. The QM makes it possible to measure the thickness of
the growing films starting from 0.2~monolayer and to use small deposition
rates to decrease overall heating due to the sample preparation. After
finishing preparation of a solid film on the surface of the QM, H
and D atoms were created by dissociation of HD or D$_{2}$ molecules
in situ in the solid by electrons created by an rf discharge \cite{Ahokas10}.
The discharge is obtained by applying pulsed rf power to a helical
resonator (HNMR) located close to the QM and operating at $\approx910\,\textrm{MHz}$.
An extra chamber with another RF coil attached to the sample cell
from the top was used as atomic hydrogen gas source. Atoms in the
gas phase can be stabilized if the sample cell walls are covered by
superfluid helium film. Having a possibility of filling the sample
cell with the gas of H or D provides good magnetic field markers for
the accurate measurement of the shifts of the ESR lines originating
from the atoms in solid hydrogen films. Measuring heat released in
recombination of gas phase atoms allows independent calibration of
the absolute number of spins in a sample. For preparation of our samples
we used hydrogen isotopes purchased from Linde AG, D$_{2}$ gas with
0.23\% residual concentration of HD, and HD gas with 3\% of H$_{2}$.
In order to obtain the matrix molecules in their ground rotational
state, prior to the condensation of the D$_{2}$ gas into the sample
cell we kept it overnight in the para-ortho converter \citep{SilveraH280}.
This ensures high concentration,$\sim$90\%, of ortho-deuterium in
our samples.

\section{Experimental results}

\begin{figure}
\includegraphics[width=1\columnwidth]{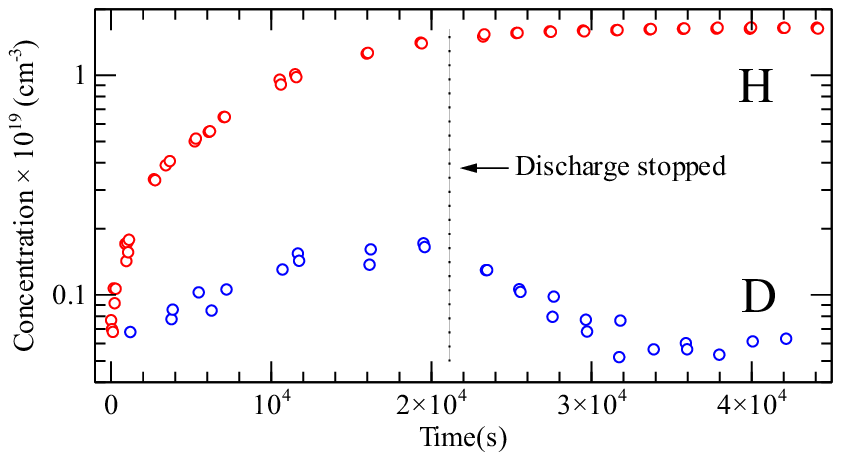} \protect \caption{Dependence of H and D atom concentrations in solid HD on time during
the process of atom accumulation by running discharge in the sample
cell and after discharge was switched off. The D concentration reaches
steady state in a time comparable with the time constant of reaction
(\ref{eq:D_HD_react}) $2\times10^{4}\,\textrm{s}$ while H concentration
continues to grow even further. The D concentration rapidly decreases
due to reaction (\ref{eq:D_HD_react}) after the discharge is switched
off. \label{fig:Accumulation}}
\end{figure}

The experimental study of the exchange reaction (\ref{eq:D_HD_react})
was carried out in two different matrices: a pure HD matrix and a
D$_{2}$ matrix with a 0.23\% HD admixture. These two systems allowed
us to study two different approaches: the first case, where the D
atoms are surrounded by 12 HD molecules, no D-atom diffusion is needed
for the reaction to proceed. In the latter case, D atoms need to diffuse
a distance of several lattice constants before they will encounter
a HD molecule and accomplish the exchange reaction.

\begin{figure}
\includegraphics[width=8cm]{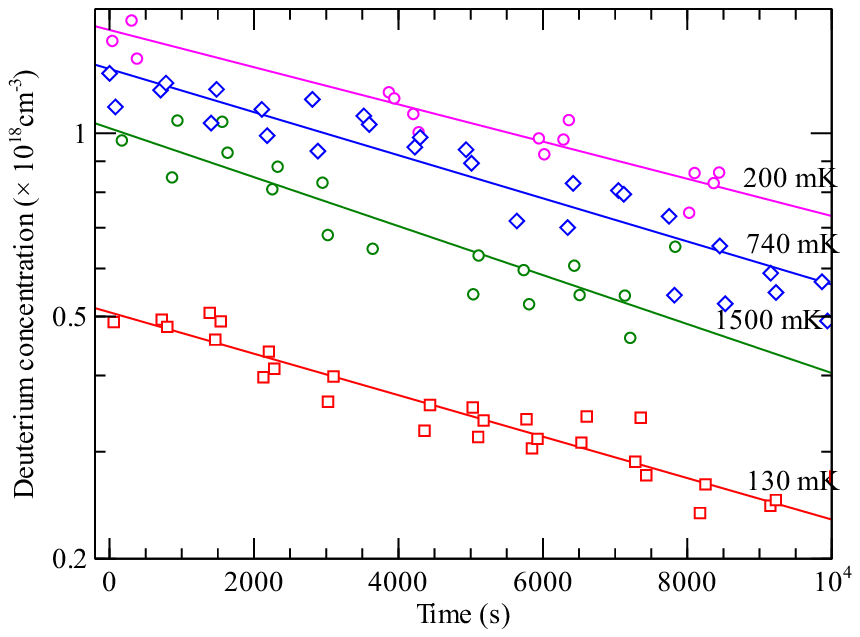} \protect\caption{Deuterium atom concentration decay in solid HD at different temperatures.
Note that the slopes of the lines are almost parallel. \label{fig:Deuterium-decays}}
\end{figure}

At the beginning of the measurements with the pure HD sample, a 200~nm
HD film was deposited onto the quartz microbalance. Then the rf discharge
was started in order to begin accumulating H and D atoms in the HD
solid. The sample cell temperature was increased to 0.7~K due to
the discharge. During accumulation, H and D atom populations were
monitored by ESR (Fig. (\ref{fig:Accumulation})) and when the D concentration
saturated, the discharge was turned off. Typically the discharge was
run for 5-12~h. The maximum H and D concentrations we achieved were
$2\times10^{19}\,\textrm{cm}^{-3}$ and $1.5\times10^{-18}\,\textrm{cm}^{-3}$,
respectively. Then the system was set to a desired temperature and
the measurements of the D atom decay began. ESR spectrum of atomic
D contains three lines separated by 78 G due to hyperfine interaction\cite{DNPPRL14}.
In thermal equilibrium the lines have equal area, and therefore for
monitoring the D concentration evolution is sufficient to record only
one line. This has been done every ten minutes until the D lines vanished
in the noise. After finishing one measurement the process of atom
accumulation was repeated. Even after several accumulations we did
not observe differences in the sample properties. Moreover exchange
reaction rates measured for different HD films were quite reproducible.
The D atom decay curves obtained at different temperatures are presented
in Fig \ref{fig:Deuterium-decays}. The rather high temperature during
discharge (0.7$\,$K) provided a problem for the lowest temperature
measurements because the system required 1-2$\,$h to cool down to
T$\approx130\,$mK during which the D concentration had already decreased
considerably.

Absolute concentrations of atoms were determined via the effect of
broadening the ESR lines due to dipole-dipole interaction between
the atoms\cite{Ahokas10}. At the high densities considered in this
work this broadening mechanism is substantially stronger than other
effects related with the nuclear dipolar moments of the matrix, and
provides reliable calibration of the ESR line area versus concentration
of atoms. Independent verification of this calibration was performed
calorimetrically, measuring the heat released in recombination of
H atoms in the gas phase. We evaluate possible systematic error in
the concentration measurement as 20\%.

\begin{figure}
\includegraphics[width=1\columnwidth]{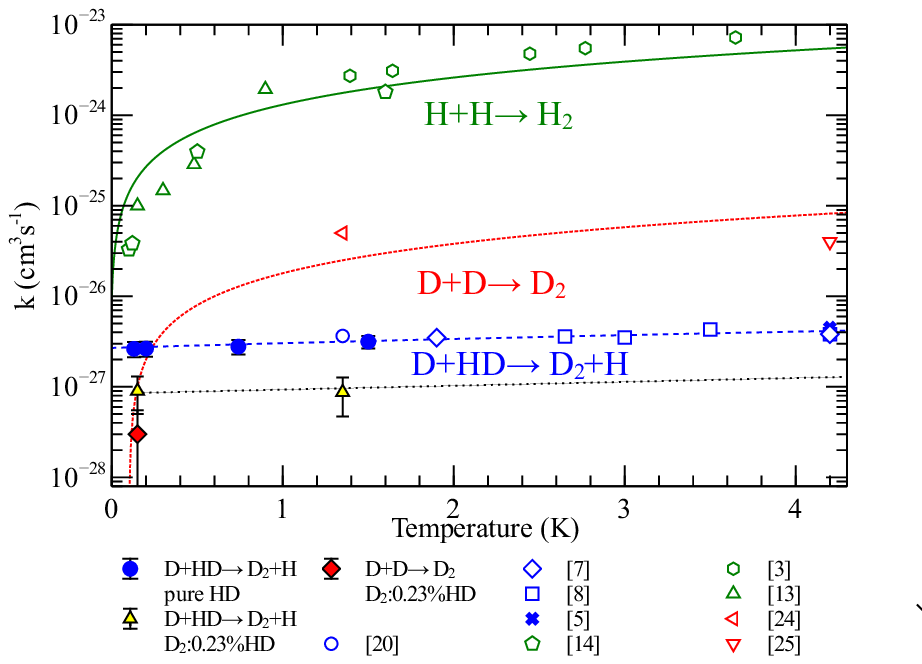} \protect\caption{The H in H$_{2}$ recombination rates (green symbols), D in D$_{2}$
recombination rates (red symbols) and the rates of the exchange reaction
(\ref{eq:D_HD_react}) (blue symbols) as a function of temperature.
The rates of the exchange reaction (\ref{eq:D_HD_react}) in D$_{2}$:0.23\%HD
matrix are shown as yellow triangles. The filled symbols correspond
to experimental data obtained in this work. The lines are shown only
as a guide. \label{fig:The-rates} }
\end{figure}

The reaction (\ref{eq:D_HD_react}) is described by the kinetic equation
\begin{equation}
\frac{d\left[\textrm{D}\right]}{dt}=-k_{ex}\left[\textrm{HD}\right]\left[\textrm{D}\right],\label{eq:HD_react_eq}
\end{equation}
where $k_{ex}$ is the rate constant. Because $\left[\textrm{D}\right]\ll\left[\textrm{HD}\right]$,
$\left[\textrm{HD}\right]$ can be considered constant and eq. (\ref{eq:HD_react_eq})
becomes a pseudo first-order reaction. Then slopes of the linear fits
of semi-log plot in Fig. \ref{fig:Deuterium-decays} give the value
of $k\left[\textrm{HD}\right]$ where we used the density of HD molecules
equal to $0.0486\,\text{mol}^{-1}\text{cm}^{-3}$ \cite{SilveraH280}.
This measurement is free of errors in the absolute concentration calibration
because the first-order reaction rate depends only on the time constant.
Possible recombination of atoms is negligibly small since the change
of the total concentration $\left[\textrm{H}\right]+\left[\textrm{D}\right]$
was not observed within the accuracy of the measurements. Our results
for $k_{ex}$ and values of these rate constants from the literature
are presented in Fig.\ref{fig:The-rates}. Takayanagi et al. carried
out theoretical calculations of $k_{ex}$\cite{Takayanagi90}. Prior
to this work the lowest temperature measurement was done by Kiselev
et al. \cite{Kiselev02} and Bernard et al. \cite{Bernard05} who
studied the reaction (\ref{eq:D_HD_react}) in HD-D$_{2}$ nanoclusters
of impurity helium condensates at temperature of 1.35\,K. The rates
obtained in these works are in a good agreement with our results.

One may expect that electrons and ions resulting from the discharge
may also be trapped in our solid samples. Sometimes we observed an
extra line in the center of ESR spectrum attributed to the trapped
electrons \cite{Electrons_QFS}. Presence of ions was also observed
in the studies of HD samples irradiated by $\gamma-$and $x$-rays
\cite{MiyazakiIons95} by Kumada et al. Although such impurities may
distort the lattice and severely influence its properties, we do not
think that this has any importance in the measurements of the exchange
reaction in pure HD, since no any diffusion of reagents is required
for the reaction. We have not found differences in the H or D atom
properties in samples which exhibit an ESR line at the free electron
resonance \cite{Electrons_QFS}.

\begin{figure}
\includegraphics[width=1\columnwidth]{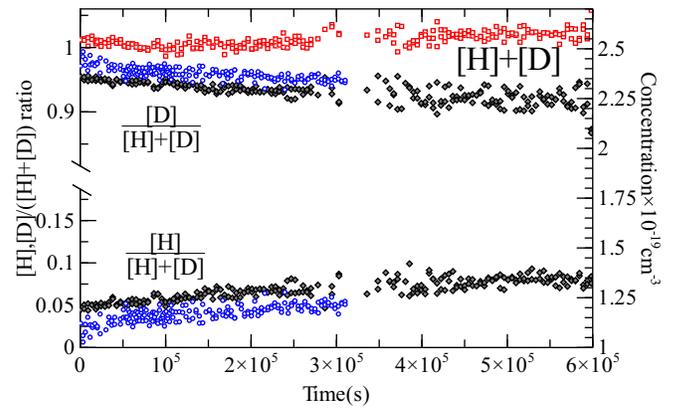} \protect\caption{The evolution of H and D atom fractions in a D$_{2}$ matrix with
a 0.23\% HD admixture as a result of reaction (\ref{eq:D_HD_react})
at temperature 150$\,$mK (grey diamonds) and 1.35$\,$K (blue circles).
The total concentration of H and D atoms measured at 150$\,$mK is
shown by red squares. The {[}H{]},{[}D{]}/({[}H{]}+{[}D{]}) axis is
shown broken to emphasize the growth of the H atom fraction. \label{fig:H_D_density_023percent}}
\end{figure}

In a second part of our work we studied the exchange reactions in
D$_{2}$ sample prepared from the pure D$_{2}$ gas, the most pure
commercially available: with an HD admixture of 0.23\% ($6.9\times10^{19}\,\textrm{cm}^{-3}$)
as specified by the manufacturer (Linde AG). It turned out that due
to the exchange reaction (\ref{eq:D_HD_react}) even this tiny amount
of HD results in a rather large concentrations of H atoms in the matrix.

A 200$\,$nm D$_{2}$ film was condensed directly from the room temperature
reservoir similar to that described before. The measurement was started
after accumulating atoms in the film by running a discharge in the
sample cell for two days and subsequently cooling to T=150$\,$mK
(Fig.\ref{fig:H_D_density_023percent}). The concentrations of D and
H atoms approached 2.3$\times$10$^{19}$cm$^{-3}$ and 2.1$\times$10$^{18}$cm$^{-3}$
after $\sim$200$\,$h of measurement. The H:D ratio is much greater
than what one should expect from direct dissociation of H and D atoms
during discharge and assuming that no exchange reaction was taking
place. Based on the gas composition the D:H ratio should be about
1700:1. The total H and D atom concentration remained constant in
the course of measurement when discharge was kept off which means
that neither H nor D recombination was taking place. \\
 The concentrations of the reactants D and HD in the D$_{2}$+0.23\%
HD sample were of the same order during the measurement. Therefore,
for determination of the reaction rates we need to solve the equation
(\ref{eq:HD_react_eq}) assuming true second order process. The solution
can be presented in a form 
\begin{equation}
\frac{1}{[\mbox{HD}]_{0}-[\mbox{D}]_{0}}\cdot ln\frac{[\mbox{HD}][\mbox{D}]_{0}}{[\mbox{D}][\mbox{HD}]_{0}}=k_{ex}t
\end{equation}
where {[}HD{]}$_{0}$ and {[}D{]}$_{0}$ are the concentrations at
the beginning of measurement while {[}HD{]} and {[}D{]} are the time
dependent concentrations of the reactants during the course of measurement.
Concentration of D atoms we determined from the ESR lines area using
the absolute density calibration as described above. The values of
{[}HD{]} were estimated indirectly from the {[}H{]} growth as {[}HD{]}={[}HD{]}$_{0}$-{[}H{]},
assuming that the exchange reaction is the only process leading to
the decrease of {[}HD{]}. Using experimental data in Fig.\ref{fig:H_D_density_023percent}
we determined the effective exchange reaction rate at 150 mK $k_{exD_{2}}=9(4)\times10^{-28}$
cm$^{3}$s$^{-1}$. We repeated the described above measurement of
the exchange reaction at temperature of 1.35 K and observed practically
same time dependence, as can be seen in Fig. (\ref{fig:H_D_density_023percent}).
Unfortunately, the long-term stability of our ESR spectrometer degraded
at temperatures above 1 K, and was not sufficient for accurate monitoring
of the H and D densities. Therefore, in Fig.\ref{fig:H_D_density_023percent}
we plotted the fractional ratios of densities {[}H{]}/({[}H{]}+{[}D{]})
and {[}D{]}/({[}H{]}+{[}D{]}), which were not sensitive to the slow
drifts of the spectrometer sensitivity and better represent the changes
caused by the exchange reactions. Using the 1.35 K data in Fig. \ref{fig:H_D_density_023percent}
we extracted the rate constant $k_{exD_{2}}\approx8\times10^{-28}$
cm$^{3}$s$^{-1}$ very close to the value obtained at 150 mK. Such
behaviour indicates that the exchange reaction proceeded with the
same rate having very weak or no temperature dependence. We attempted
similar measurements at higher temperatures of 2.5 and 3.5 K. In addition
to the conversion of D into H we also observed a decrease of the total
density of atoms caused by their recombination. Since the recombination
includes several channels including H+H, H+D, and D+D processes, and
also because of the increased drifts at higher temperatures, we have
not attempted to extract any quantitative data from these measurements.
Such work requires optimization of the experimental setup for operation
at higher temperatures and may be performed in the future.

\section{Discussion}

In this work we performed measurements of the hydrogen isotope exchange
reactions in the matrices of pure HD and D$_{2}$. Other well known
reactions which occur in these systems are the reactions of recombination
of atoms into molecules. These reactions are also of the second order
in density. This allows direct comparison of the rates, which we perform
in Fig.\ref{fig:The-rates} where the rate constants of the exchange
reaction (\ref{eq:D_HD_react}), the recombination rate constant of
H in H$_{2}$, and D in D$_{2}$ are presented for the temperature
range 0.15-4~K. The plot includes results of the present work as
well as a compilation of our previous results and results of other
groups. The scarcity of the reactants ({[}H{]} in H$_{2}$ $\ll$
{[}HD{]} in pure HD) requires a much longer time to detect the H atom
decay although the reaction rate turns out to be greater than that
of the exchange reaction (\ref{eq:D_HD_react}) (Fig.\ref{fig:The-rates}).

The measurements in pure HD demonstrate that the exchange reaction
takes place at the lowest temperature at the same rate as at 1.5~K
without any clear temperature dependence. Such behaviour is expected
for tunnelling reaction and means that the chemical reactions of this
type may proceed down to absolute zero of temperature. In this case
we studied the pure exchange reaction (\ref{eq:D_HD_react}), because
each D atom is surrounded by 12 HD molecules needed for the reaction,
and diffusion stage is not required. The total H and D atom concentration
in the HD samples we studied was about 3$\times$10$^{19}\,\textrm{cm}^{-3}$
which corresponds to the mean distance of 3~nm or 8 lattice constants
between them. The H-D or D-D interaction at such distances becomes
negligibly small and cannot influence the rate of the exchange reaction
(\ref{eq:D_HD_react}) \cite{Kumada02}.

In the samples of D$_{2}$ with small concentration of HD dissociation
of atoms by the discharge produces mainly D atoms which on the average
are located quite far away from HD impurities. We may neglect production
of H and D atoms by the direct dissociation of HD, because its concentration
in the matrix is much smaller than D$_{2}$. Since each D atom has
12 neighbours, the probability that one of the neighbours is HD may
be estimated as $1-0.9977^{12}\sim2.7\%$. However, as we can see
from Fig. \ref{fig:H_D_density_023percent}, almost 10\% of D atoms
were converted to H which provides evidence for D atom diffusion.
The complete sequence of the D-to-H conversion in this case includes
the diffusion process $\textrm{D}+\textrm{D}_{2}\rightarrow\textrm{D}_{2}+\textrm{D}$
and the exchange reaction (\ref{eq:D_HD_react}) itself. This provides
an opposite limit to the described above experiments with pure HD
where diffusion is not required at all.

Absence of temperature dependence in our data on the exchange reaction
rate in the D$_{2}$ matrix indicate that the process is not diffusion
limited, and the reaction itself is the limiting stage. Then, comparing
the absolute values of the rate constant we note that in the HD matrix
it is only a factor of 3 larger than in D$_{2}$. Since the diffusion
stage could further slow down the effective rate of the exchange reaction
in the D$_{2}$ matrix, the difference in the tunnelling rates of
the reaction may be even smaller. Considering that in hcp lattice
of pure HD each D impurity is surrounded by 12 neighbours, one may
expect that in D$_{2}$ matrix the exchange reaction should be at
least 12 times slower. We would like to point out also on a substantial
difference in the initial and final states of the reactants in these
two matrices. For the exchange reaction (\ref{eq:D_HD_react}) in
pure HD the initial state has a single crystal defect, which can be
represented in 1 dimensional case as ...HD-HD-\textbf{D}-HD-HD...
After the act of the exchange reaction we obtain two defects: ...HD-HD-\textbf{D$_{2}$}-\textbf{H}-HD...,
which in total create much larger distortion of the lattice than in
the initial state. On the contrary, in the D$_{2}$ matrix we have
initially ...D$_{2}$-D$_{2}$-\textbf{D}-\textbf{HD}-D$_{2}$...,
which after the exchange is converted to: ...D$_{2}$-D$_{2}$-\textbf{H}-D$_{2}$-D$_{2}$...
Now the situation is opposite, two lattice defects are converted into
one. Although, the mass is conserved in the exchange reactions, the
volume of the lattice distortion is much larger in the case if two
neighbouring defects, which may lead to a larger difference in the
energies of the initial and final states for the reaction in the HD
matrix. We consider this as a possible explanation for the observed
difference in the exchange reaction probabilities in HD and D$_{2}$
matrices.

The recombination of H atoms in H$_{2}$ solids also includes two
stages: approaching of H atoms towards each other by a distance of
one lattice constant by a series of tunnelling reactions $\textrm{H}+\textrm{H}_{2}\rightarrow\textrm{H}_{2}+\textrm{H}$
followed by the formation of a H$_{2}$ molecule by the recombination
reaction $\textrm{H}+\textrm{H}\rightarrow\textrm{H}_{2}$ with a
transfer of recombination energy to the lattice. For spin-polarized
atoms in the lattice there appears to be several fast depolarization
processes which lead to a flip of the electron spin of one of the
atoms and subsequent recombination. It has been shown that the rate
of this process does not depend on magnetic field at temperatures
above 1.3 K \cite{Ivliev85}, and the diffusion stage is always the
limiting stage of the recombination.

Hydrogen migration in a perfect H$_{2}$ crystal proceeds via resonant
tunnelling when the initial and final states of a tunnelling event
coincide. The presence of H atoms (or other crystal imperfections)
in the lattice causes the mismatch of the initial and final levels
which greatly reduces the tunnelling rate. The mismatch increases
when H atoms approach each other and finally it prevents them from
occupying the neighbouring lattice sites where they are able to recombine.
The level mismatch can be overcome by phonon assistance via a single-phonon
direct process or a two-phonon Raman process, which lead to a strong
temperature dependence of the recombination rate.

Data for the recombination of H atoms in solid H$_{2}$ summarized
in Fig. \ref{fig:The-rates} showed a strong temperature dependence
with two distinct ranges: 1.3-4~K where it depends linearly on temperature
and below 1$\,$K where the rate constant decreases substantially
and become immeasurably small at 150$\,$mK\cite{HinH2_06}. Recombination
becomes strongly inhibited at the lowest temperatures, about 150$\,$mK
where no phonons are available to compensate for the level mismatch.
Ahokas et al. showed that H-atom recombination can be initiated at
temperatures below 500~mK if phonons are injected into the system
\cite{Ahokas10} as a result of recombination of the gas-phase H$\downarrow$
atoms on the H$_{2}$ film surface.

Recombination of atomic deuterium in solid D$_{2}$ has similar dependence
on temperature, and is also diffusion limited. The absolute values
of the rate constants are two orders of magnitude smaller, and approaching
to 1 K one cannot detect any recombination making its measurement
impossible. The upper limit estimate for the D atom recombination
constant at 150$\,$mK found from Fig.\ref{fig:H_D_density_023percent}
is equal to $k_{rD_{2}}\simeq3(2)\times10^{-28}\mbox{cm}^{-3}.$ The
value of $k_{rD_{2}}$ at T=150$\,$mK is much smaller than that measured
at 4.2$\,$K and 1.5$\,$K reported earlier (Fig.\ref{fig:The-rates})\cite{RusChemRev2007},
\cite{Iskovskikh86} which gives evidence for a strong temperature
dependence of the recombination constant of D atoms below 1.5$\,$K,
similar to that observed for H in H$_{2}$ below 1$\,$K\cite{Ahokas10}.
Heavier D atoms in a D$_{2}$ crystal have smaller tunneling probability
than for H in H$_{2}$ and diffuse much slower.

As we see, the diffusion rate of the reactants of the exchange reaction
(\ref{eq:D_HD_react}) in the D$_{2}$ matrix is substantially faster
than the diffusion of two D atoms approaching each other for recombination.
We suggest two possible reasons for that. First, there is a smaller
mass difference $\Delta m$ between a HD molecule and hosting D$_{2}$
molecules relative to the mass $M$ of the host molecule: $\Delta m/M=1/4$
compared to that of a $\Delta m/M=1/2$ for D in D$_{2}$ (and also
for H in H$_{2}$). This implies a smaller perturbation of the periodic
potential of the matrix in the former case. Second, it is known that
the HD molecules in their ground state may rapidly move in the D$_{2}$
crystals via tunnelling exchange with ortho-D$_{2}$\cite{sullivan95}.
This process may go even faster than the diffusion of D atoms due
to the reaction (\ref{eq:D_D2_diff}). In this case, the energy level
mismatch should be overcome as well. As a result, it becomes easier
to approach each other for D atom and HD molecule than for two D atoms.
The recombination rate of D atoms in a D$_{2}$ matrix should be slower
than the rate of the exchange reaction (\ref{eq:D_HD_react}) as observed
in the present work.

The exchange reaction (\ref{eq:D_H2_react}) is expected to proceed
two orders of magnitude faster than the reaction (\ref{eq:D_HD_react})
and cannot be measured in a pure H$_{2}$ matrix with the accumulation
technique used in our experiments. Measurements of samples with a
small fraction of H$_{2}$ are diffusion limited and therefore do
not represent the true reaction rate \cite{Kumada06}. But such measurements
will yield information on the diffusion of D if the reaction (\ref{eq:D_H2_react})
remains fast at ultralow temperatures. One should be able to make
lower limit estimates of the reaction (\ref{eq:D_H2_react}) by comparing
the yields of D and H in samples with different $\textrm{D}_{2}:\textrm{H}_{2}$
mixtures. For H$_{2}$ fractions above 25\% no D lines are expected
\cite{RusChemRev2007}.

\section{Conclusions}

In conclusion, we have reported on the first experimental observation
of the isotopic exchange tunnelling reaction, $\textrm{D}+\textrm{HD}\rightarrow\textrm{D}_{2}+\textrm{H}$,
taking place in a solid matrix of hydrogen deuteride at temperatures
below 1$\,$K. The reaction rate, $k_{ex}$, was measured within the
temperature range 1.5$\,$K - 130$\,$mK where it was found to be
nearly temperature independent, $k_{ex}\approx3\times10^{-27}$cm$^{3}$s$^{-1}$.
The exchange reaction also takes place at T=150$\,$mK and 1.35$\,$K
in a D$_{2}$ matrix with 0.23\% HD admixture where D atoms need to
migrate by the distance of several lattice constants in order to participate
in the reaction. The reaction rate did not change upon raising temperature
from 150$\,$mK to 1.35$\,$K. The fact that the isotopic exchange
reaction D$+$HD$\rightarrow$D$_{2}$+H proceeds there suggests that
the exchange reaction D$+$D$_{2}$$\rightarrow$D$_{2}$$+$D which
governs D atom diffusion in a D$_{2}$ matrix is substantially faster
than the exchange reaction $\textrm{D}+\textrm{HD}\rightarrow\textrm{D}_{2}+\textrm{H}$.
Recombination of D atoms becomes suppressed due to the energy level
mismatch appearing when two atoms approach each other. The level mismatch
depends on the mass difference between the atomic impurity and the
molecule of the host lattice, and is substantially smaller for the
case of HD in the D$_{2}$ matrix.

These observations have a bearing on the rate of the exchange reaction
H$+$H$_{2}\rightarrow$H$_{2}$+H also does not depend on temperature
if H and H$_{2}$ occupy neighboring sites even though the rate of
the atomic hydrogen recombination reaction, $\textrm{H}+\textrm{H}\rightarrow\textrm{H}_{2}$,
is substantially reduced upon lowering temperature below 1~K and
it almost completely vanishes at 150$\,$mK. Therefore we anticipate
that recombination of H atoms in solid H$_{2}$ is mainly limited
by the diffusion stage, the exchange reaction $\textrm{H}+\textrm{H}_{2}\rightarrow\textrm{H}_{2}+\textrm{H}$,
which is inhibited in samples with high atomic concentration at low
temperatures where the condition of resonant tunneling becomes violated.

The experiments show strong decreasing of recombination rates of H
(D) atoms in solid H$_{2}$ (D$_{2}$) due to suppression of diffusion
of atoms at temperatures below 1$\,$K. New theoretical approaches
are needed to understand this phenomenon.

\section*{Acknowledgements}

We acknowledge the funding from the Wihuri Foundation and the Finnish
academy grants No. 258074, 260531 and 268745. This work is also supported
by NSF grant No DMR 1209255. S.S. thanks UTUGS for support.

 \bibliographystyle{apsrev4-1}
\bibliography{exchange_reaction_05_07_toarxiv}

\end{document}